\documentclass[letterpaper]{article} 
\usepackage{aaai25}  
\nocopyright 
\usepackage{times}  
\usepackage{helvet}  
\usepackage{courier}  
\usepackage[hyphens]{url}  
\usepackage{graphicx} 
\usepackage{natbib}  
\usepackage{caption} 
\usepackage{algorithm}
\usepackage{algorithmic}

\usepackage{newfloat}
\usepackage{listings}
\usepackage{lipsum}

\DeclareCaptionStyle{ruled}{labelfont=normalfont,labelsep=colon,strut=off} 
\floatstyle{ruled}
\newfloat{listing}{tb}{lst}{}
\floatname{listing}{Listing}
\usepackage{subcaption}  
\usepackage[export]{adjustbox}   

\title{Breakable Machine:  A K–12 Classroom Game for Transformative AI Literacy Through Spoofing and eXplainable AI (XAI)
}
\author{
    Olli Hilke\textsuperscript{\rm 1},
    Nicolas Pope\textsuperscript{\rm 1},
    Juho Kahila\textsuperscript{\rm 2},
    Henriikka Vartiainen\textsuperscript{\rm 2},
    Teemu Roos\textsuperscript{\rm 3},
    Tuomo Parkki\textsuperscript{\rm 4},
    Matti Tedre\textsuperscript{\rm 1}
}
\affiliations{
    \textsuperscript{\rm 1}University of Eastern Finland, School of Computing, Joensuu, Finland\\
    \textsuperscript{\rm 2}University of Eastern Finland, Applied Educational Science and Teacher Education, Finland\\
    \textsuperscript{\rm 3}University of Helsinki, Department of Computer Science, Helsinki, Finland\\
    \textsuperscript{\rm 4}Joensuu Lyseo School, Finland\\    
    \{olli.hilke, npope, juho.kahila, henriikka.vartiainen\}@uef.fi, teemu.roos@helsinki.fi, tuomo.parkki@edu.joensuu.fi
}

\begin{document}

\maketitle

\begin{abstract}
This paper, submitted to the special track on resources for teaching AI in K--12, presents an eXplainable AI (XAI)-based classroom game ``Breakable Machine'' for teaching critical, transformative AI literacy through adversarial play and interrogation of AI systems. Designed for learners aged 10–15, the game invites students to spoof an image classifier by manipulating their appearance or environment in order to trigger high-confidence misclassifications. Rather than focusing on building AI models, this activity centers on breaking them---exposing their brittleness, bias, and vulnerability through hands-on, embodied experimentation. The game includes an XAI view to help students visualize feature saliency, revealing how models attend to specific visual cues.  A shared classroom leaderboard fosters collaborative inquiry and comparison of strategies, turning the classroom into a site for collective sensemaking. This approach reframes AI education by treating model failure and misclassification not as problems to be debugged, but as pedagogically rich opportunities to interrogate AI as a sociotechnical system.  In doing so, the game supports students in developing data agency, ethical awareness, and a critical stance toward AI systems increasingly embedded in everyday life. The game and its source code are freely available.
\end{abstract}

%

\begin{links}
    \link{App}{[omitted-for-review]}
    \link{Code}{[omitted-for-review]}
\end{links}

\section{Introduction}

The rapid rise of artificial intelligence (AI) has fueled interest in bringing modern machine learning (ML) concepts into K–12 education, with the trend increasingly extending to pre- and in-service teacher training \cite{sanusi22,eguchi25,grover24,rizvi23,martins22,shapiro19,long23,olari24,zhang23}. This interest in AI education---focused not on how to use AI, but on understanding how AI systems work---reflects recognition of the influence of AI-driven technology on the future of work, societal structures, and everyday life \cite[e.g.,][]{grover24,hoper23,sentance22}. 

Current K–12 AI education efforts address AI concepts and techniques at different levels of abstraction---such as classifiers, pattern recognition, or neural networks---as well as broader themes, including AI ethics, curriculum integration, teacher professional development, and AI literacy \cite{heintz21,grover24,druga22,black24,lee22}.  Some are focused on data---including topics such as data collection, curation, storage, and processing, with the overarching aim of cultivating students’ data awareness and fostering data agency \cite{olari24b,hoper23,vartiainen24c,morales24,fagerlund25}.  Others are focused on AI techniques---including topics such as neural networks, classifiers, reinforcement learning, language models, word embeddings, and more \cite{touretzky25,jatzlau19,kahila24,morales25,wiatrek25}.  Many education efforts map their learning outcomes on one or more popular AI content frameworks, such as Long and Magerko's (\citeyear{long20}) competency framework, the AI4K12 ``Five Big Ideas in AI'' framework \cite{touretzky23}, or UNESCO's competency framework \cite{miao24}.  

Some AI education initiatives have advocated for transformative agency, which goes beyond awareness and literacy by focusing on the learners' capacity to not just critically understand and challenge, but also reshape AI systems and the societal structures in which they are embedded \cite{iivari24b,vartiainen25e,veldhuis25,stetsenko19}.  Transformative agency emphasizes the development of ethical-political awareness, volition, and collective action to resist harmful data practices and reimagine more just alternatives---as well as individual and collective efforts to break away from existing practices, take initiative, and design new futures.

\def\minipicturesize{0.9} 
\def\miniminipicturesize{0.235} 
\def\miniminiscreenshotsize{0.705} 
\def\miniminiteacherviewsize{0.325} 

\begin{figure*}[t]
    \centering

    \begin{subfigure}[t]{\miniminipicturesize\textwidth}
        \centering
        \captionsetup{justification=centering}
        \includegraphics[height=200px, frame]{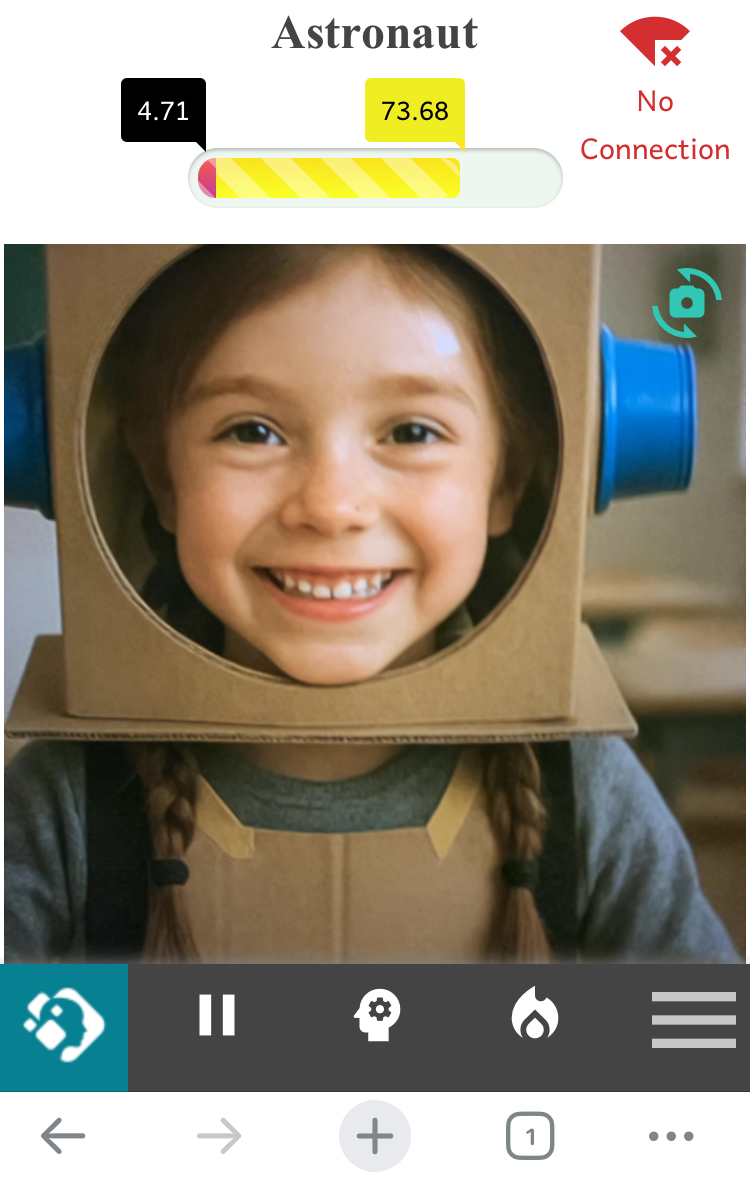}
        \caption{The game app}
        \label{fig:spoofingapp}
    \end{subfigure}
    ~ ~
    \begin{subfigure}[t]{\miniminiscreenshotsize\textwidth}
        \centering
        \captionsetup{justification=centering}
        \includegraphics[height=200px, frame]{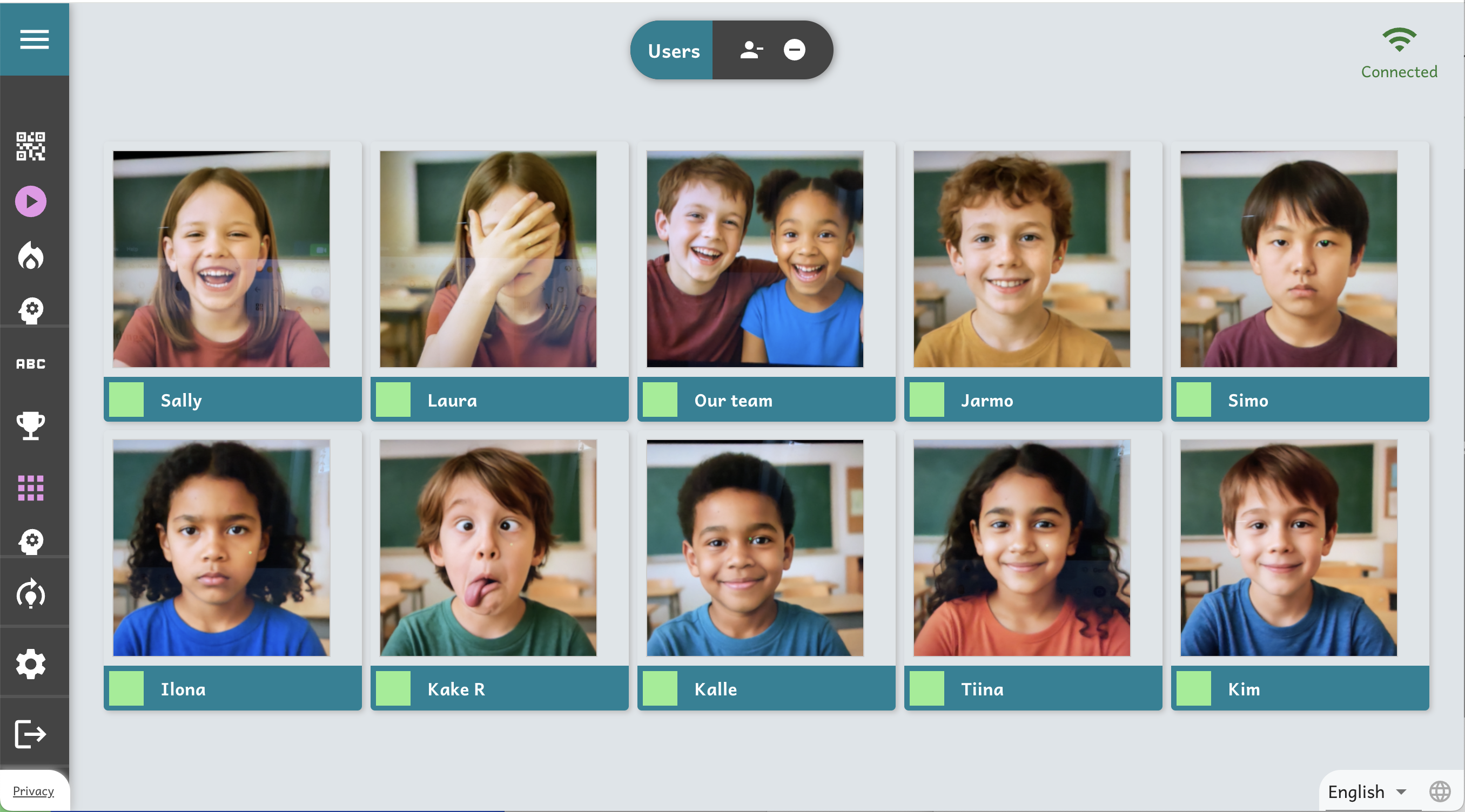}
        \caption{A view of all students' profile pictures, shown on the classroom projector.}
        \label{fig:usergrid}
    \end{subfigure}
    \caption{The game players work alone or in small groups, using a mobile phone with constant camera feed. (Fig.\ \ref{fig:spoofingapp}).  Players try to find props, backgrounds, clothing, and other means to increase the classifier confidence (on label ``astronaut'' in this case.)  At the same time, all player profile pictures and names are shown on the classroom projector (Fig.\ \ref{fig:usergrid}).}
\end{figure*}
AI education interventions employ a wide range of modalities.  Many tools allow students to explore AI techniques using a range of gadgets, apps, and games \cite{wiatrek25,touretzky25,kahila24b,cardenas23}.  AI auditing activities focus on interrogating existing systems \cite{morales24b}, and unplugged activities do not require computers \cite{lindner19}.  One of the most popular approaches involves training one's own models---for instance, with Google Teachable Machine, Generation AI Teachable Machine, MIT's Personal Image Classifier, or IBM’s Machine Learning for Kids \cite{carney20,pope25b,tang19b,lane21}.

A key learning moment in model training arises when the system behaves unexpectedly: the user encounters an ``expectation mismatch'' between the model's output and their own expectation of what it should be \cite{dhanorkar21}.  Such unexpected or incorrect predictions prompt the learner to dig deeper into the training data and process \cite{tedre21b,olari24b}.  While most prior educational tools have focused on training AI systems, some classroom tools---such as ``Erase Your Face''\footnote{https://youthradio.github.io/erase-your-face/}  \cite{abidoun20}---have instead focused on how to break image classifiers.

The tool presented in this paper elaborates the ``breaking AI systems'' theme by offering a playful, conceptually rich learning experience that highlights the concepts of trust, robustness, and accountability in AI.  To better support students' transformative AI agency, we have developed an eXplainable AI (XAI)-driven classroom game that teaches children about the kinds of failures typical of many AI systems through engaging in spoofing an AI system; i.e., carrying out an adversarial attack against an image recognition system.

The \textit{Breakable Machine} game aims to facilitate the development of a critical, transformative stance toward AI by enabling learners to take control over AI systems through hands-on experimentation, guided reflection, and collective sense-making.  In the game, students are challenged to manipulate their appearance---such as what they wear or what is seen in the background---to ``trick'' an image recognition system into producing a high-confidence classification for a given label (e.g., ``doctor''). All students in the class attempt the challenge in parallel, and the system displays each student’s highest-confidence result on a shared leaderboard projected on the teacher's screen.

This technical, non-empirical paper describes the design and pedagogical foundations of the game, illustrated with screenshots and classroom use scenarios. As empirical evaluation of learning outcomes is scheduled for future work, this paper outlines the learning objectives and theoretical rationale for this approach to teaching critical AI literacy.


\section{Description of the Resource}

\begin{figure*}[!t]
     \centering

    \begin{subfigure}[t]{\miniminiteacherviewsize\textwidth}
        \centering
        \captionsetup{justification=centering}
        \includegraphics[height=240px,frame]{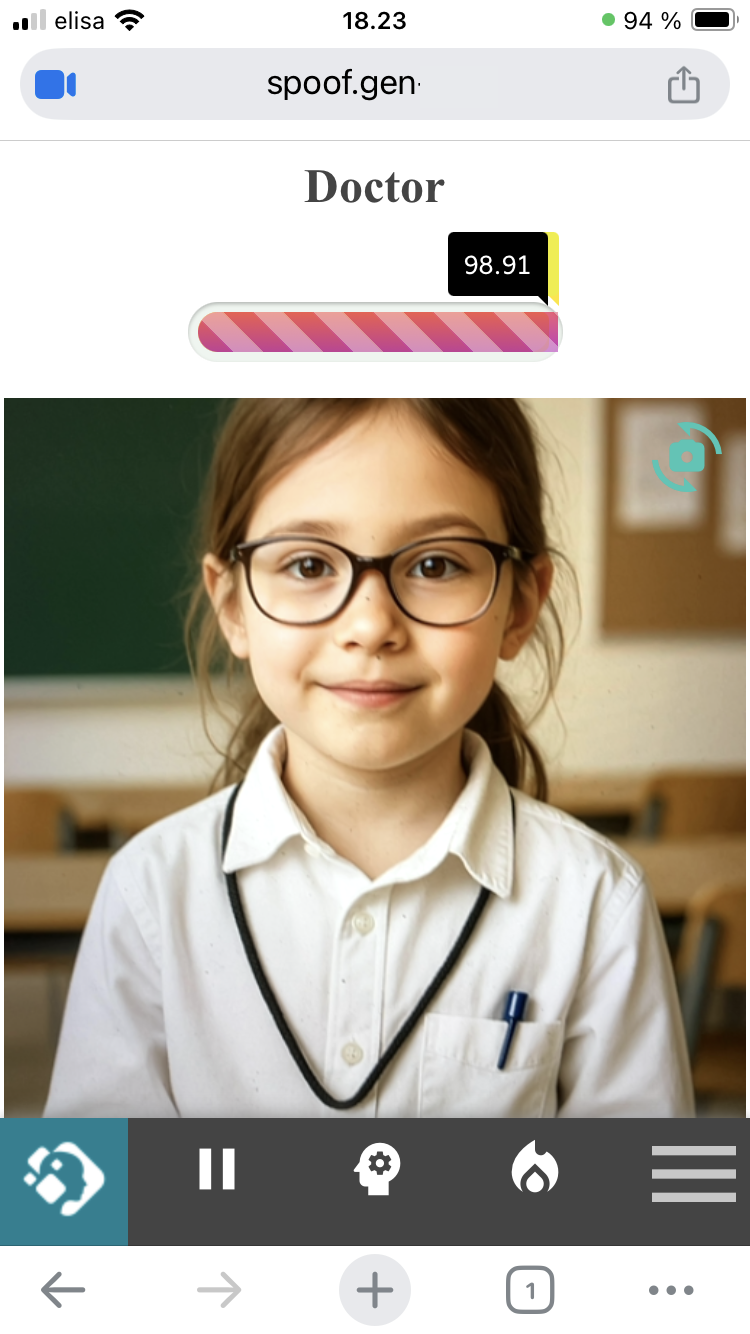}
        \caption{Spoofing app interface.}
        \label{fig:app_interface}
    \end{subfigure}%
    ~ 
    \begin{subfigure}[t]{\miniminiteacherviewsize\textwidth}
        \centering
        \captionsetup{justification=centering}
        \includegraphics[height=240px,frame]{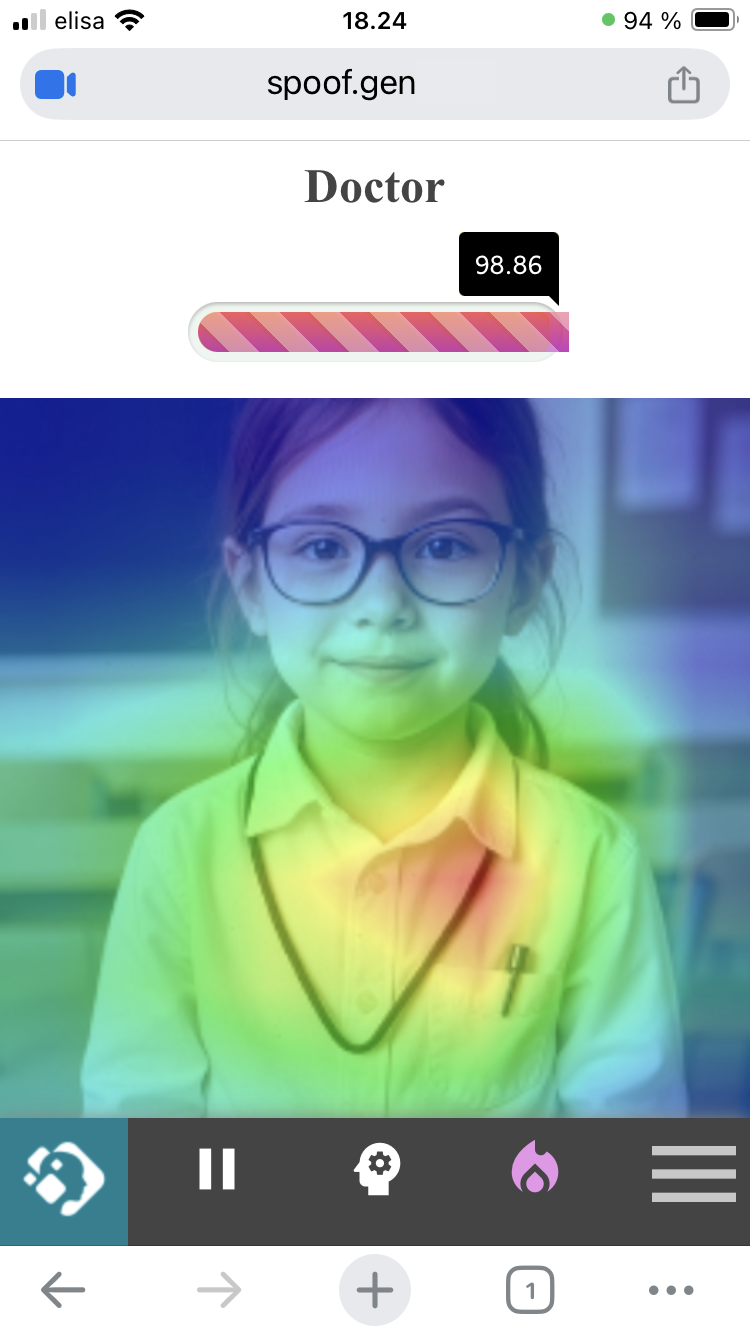}
        \caption{Heat map of salient image regions.}
        \label{fig:app_heatmap}
    \end{subfigure}
    ~ 
    \begin{subfigure}[t]{\miniminiteacherviewsize\textwidth}
        \centering
        \captionsetup{justification=centering}
        \includegraphics[height=240px,frame]{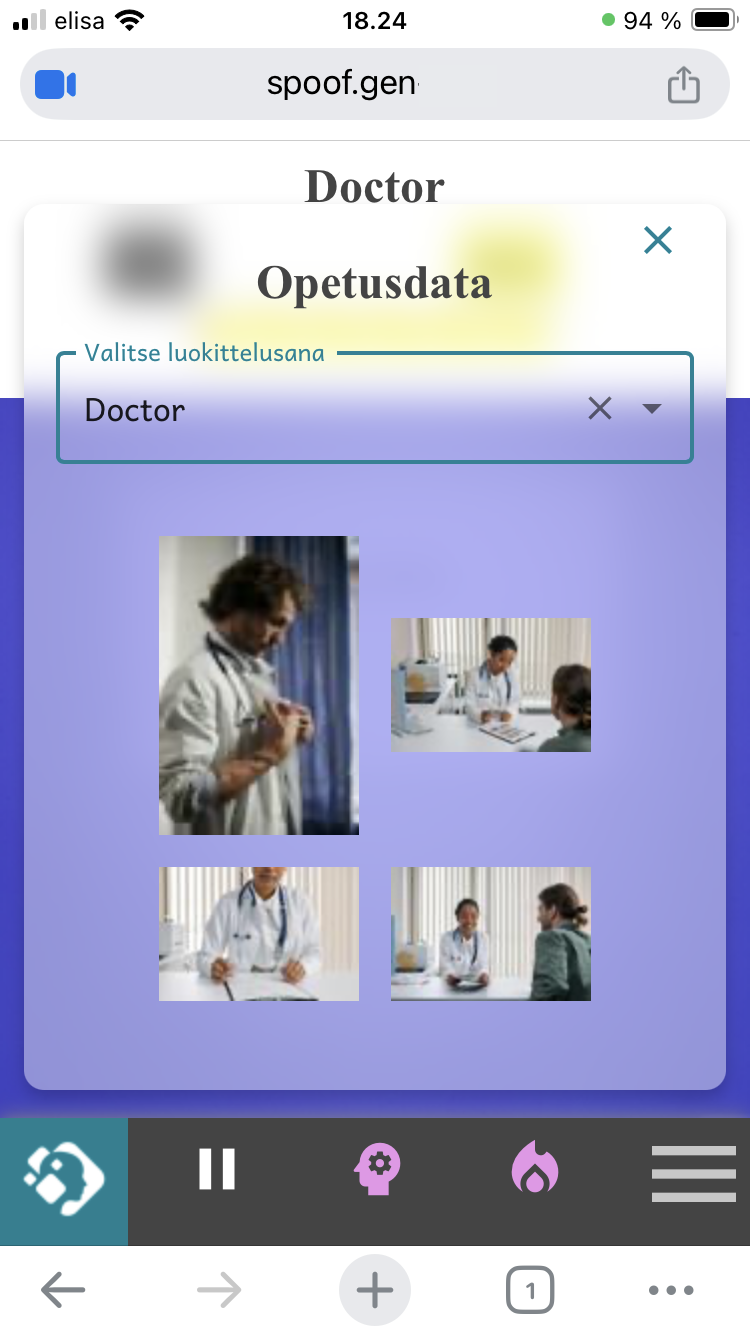}
        \caption{Examining the training data.}
        \label{fig:app_trainingdata}
    \end{subfigure}%
    \caption{Three app views.  The basic interface shows the camera feed and classification confidence for label ``Doctor'' (Fig.\ \ref{fig:app_interface}).  Users can examine an XAI heat map that highlights what image areas contribute the most to the classification result (in Fig.\ \ref{fig:app_heatmap} white shirt, shoe string, and pen in pocket).  Users can explore the training data set to get hints on spoofing the system (Fig.\ \ref{fig:app_trainingdata}).
    }
    \label{fig:recommendation_systems}
\end{figure*}

\begin{figure*}[t]
    \centering
    \captionsetup{justification=centering}
    \includegraphics[height=270px, frame]{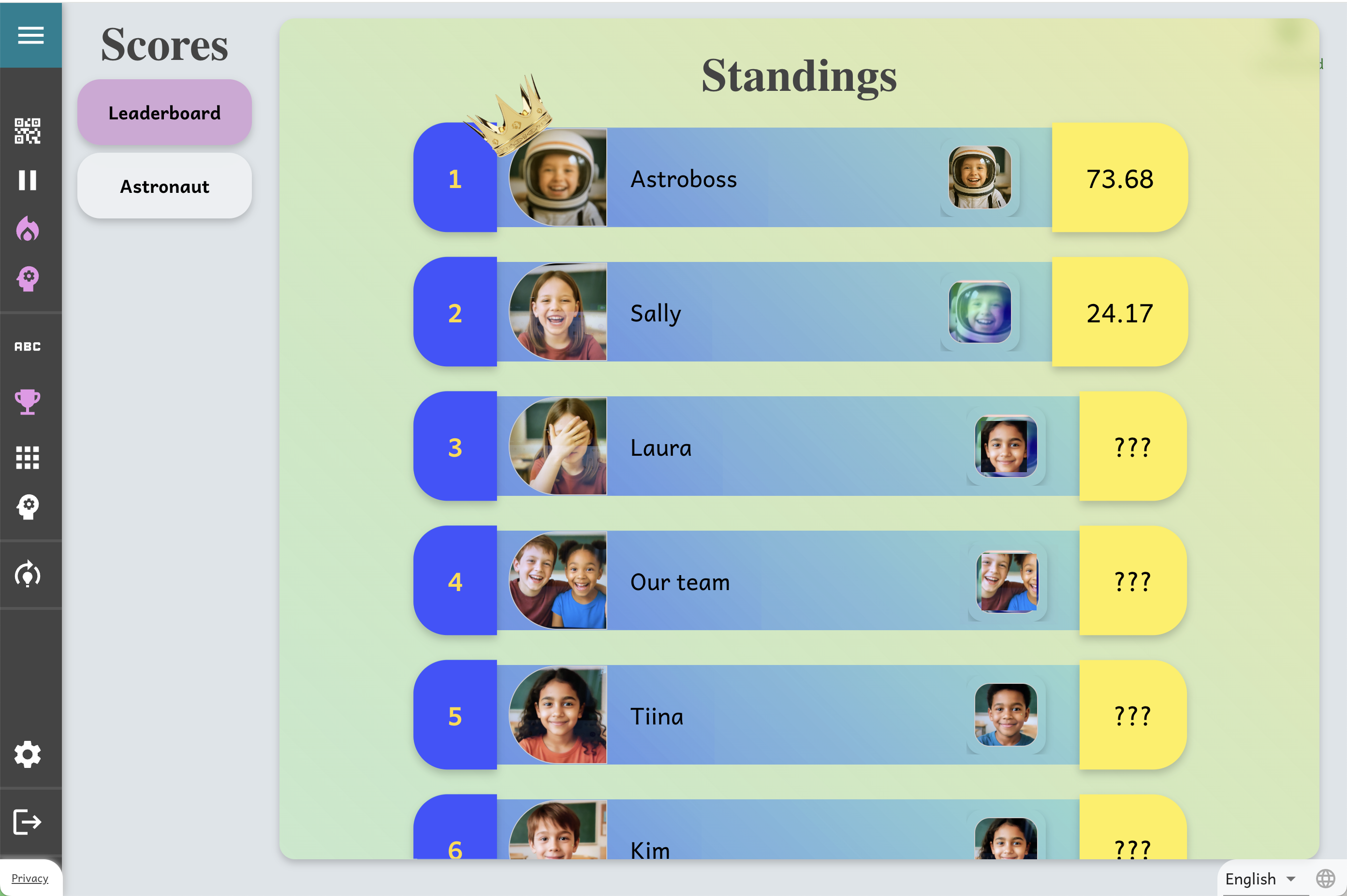}
    \caption{A view of top scoring players, with confidence score and a thumbnail of the image that scored it, shown on the classroom projector.  The confidence scores can be completely hidden or shown only for the \emph{n} highest scoring players (currently just two top players).  This view is aimed at giving players hints from others' high scoring images.}
    \label{fig:leaderboard}
\end{figure*}

\paragraph{Gameplay.}  
The teacher starts the classroom app, shown on the classroom projector,  providing students with a QR code to join the game on their own devices.  The teacher then selects a challenge for the students  (e.g., ``doctor'', ``bear'').  Students try to modify their appearance or surroundings---such as by adding eyeglasses, changing the lighting, or holding up objects---in front of their webcam to try to make the system classify them with high confidence as the target label (Fig.\ \ref{fig:spoofingapp}).  A live leaderboard tracks each student's highest-confidence result, providing hints to other students, thus turning the classroom into a collaborative inquiry space for probing why the system responds as it does (Fig.\ \ref{fig:leaderboard}).

\paragraph{Game Interfaces.}  
The game consists of two browser apps: One running on the teacher's device, projected for the entire classroom (Fig.\ \ref{fig:usergrid}), and another running on students' mobile devices (Fig.\ \ref{fig:spoofingapp}). The profile pictures here are replaced with AI-generated images for privacy. The teacher app has controls for pausing the game on all devices, changing the current challenge for all or for selected students, enabling heat map visualization on all devices, and unlocking access to the training data set on all devices.  The teacher app also has views for showing all users, displaying the scoreboard and students' top-scoring images, and exploring the training dataset together.  The student app can switch between live play, heat map exploration of the current camera feed, and browsing the training dataset.

\paragraph{Setup And Resources Required.}
The game is fully browser-based and works on browsers that support WebRTC data channels and ES11 (Chrome 56+, Firefox 44+, Safari 15.4+ and Edge 79+).  It has been tested with devices ranging from laptops (Chromebooks, Windows, Mac and Linux) to mobile phones and tablets (Android, iOS).  The classifier used is MobileNet V2, fine-tuned separately for each classification task.

\paragraph{Privacy and Security.}
Designed with children's data privacy in mind, the game adheres to EU's GDPR regulations.  It does not collect, send, or store any identifiable data outside the classroom.  All data are kept local, shared only with the teacher's device for the duration of the session, and automatically deleted when the session ends.  The only external data retrieved are the app itself and the image and label dataset.  The tool uses WebRTC-based peer-to-peer communication, and requires only local network connectivity.

\paragraph{eXplainable AI.}
Similar to some recent educational XAI techniques \cite[e.g.,][]{wang21,melsion21}, our approach uses Class Activation Maps (CAMs), which provide a visual heatmap indicating which areas of an input image contributed the most to the classification result \cite{zhou16}.  CAMs are generated from the final convolution layer that provides a classification result for each of the 7 x 7 positions.

\section{Intended Learning Outcomes}

\paragraph{Target age groups and context of learning.}

The AI spoofing education game is designed for grade 4-9 classrooms (learners aged 10–15 years) but it is suitable for all novice learners. It is particularly suitable for classrooms that explore AI concepts through hands-on experimentation, play, and critical inquiry. The game requires each player to have a mobile device with a camera, an Internet browser, and Internet or local network connectivity.  It includes a structured classroom activity to guide the learning process.

\paragraph{Prerequisites.}  Students should be familiar with the basic workings of image classification systems, including foundational concepts such as training data, classification, labels, prediction, and confidence.  Tools such as Generation AI Teachable Machine\footnote{https://tm.gen-ai.fi/}, Personal Image Classifier\footnote{https://classifier.appinventor.mit.edu/}, Machine Learning for Kids\footnote{https://machinelearningforkids.co.uk/}, or Google Teachable Machine\footnote{https://teachablemachine.withgoogle.com/} can serve as effective starting points. 

\paragraph{Pedagogical approach.}

Because the game requires familiarity with core concepts of image classification, it is pedagogically appropriate to embed it within a larger learning project rather than treating it as a short, standalone exercise. For example, it can be integrated into the CEDE pedagogical model \cite{vartiainen25e} that positions children as designers and knowledge creators in AI. In such projects, students develop their conceptual understanding, creative abilities, and critical thinking by designing and making their own ML-driven mobile phone apps using the Generation AI Teachable Machine \cite{kahila24,pope25b}. At the end of the GenAI project, the children reflect on the potential risks and harms of AI through their own app design process and are scaffolded to apply their evolving AI understanding to critical analysis and discussions of ethical and societal issues, such as algorithmic bias \cite{vartiainen24b}.

While tools like the Teachable Machine foster creative learning and collaborative inquiry, the spoofing game extends and deepens these knowledge-creative projects by adding new affordances for the \textit{ethical and social reflection} phase that is central to CEDE-based learning projects \cite{vartiainen25e}.  It responds to recent calls for AI literacy education to go beyond technical understanding and foster critical thinking, ethical awareness, and an ability to interrogate AI systems as sociotechnical constructs \cite{morales23,morales24}. Rather than presenting AI as a reliable or neutral decision-maker, the game positions AI as a fallible, human-made system, which can be intentionally probed, manipulated, and tricked.

In the game, advances in conceptual understanding through critical inquiry of image recognition systems are supported through hands-on learning experiences. Students are guided to test and construct their conceptions of image recognition by making observations and experiments, and by explaining their conclusions with systematic evidence \cite{aleknaviciute23,osterhaus21}. This approach makes abstract AI concepts more accessible, concrete and visible \cite{smetana12,trundle10}, and situates them in the context of real-world ethical issues \cite{vartiainen25e}. Furthermore, the game allows students to challenge and test their prior beliefs and conceptions by visualizing objects and processes that are normally hidden and from perception and manipulation \cite{trundle10,smetana12}. Importantly, it also becomes a medium that externalizes learners' evolving ideas and lines of reasoning by making them tangible and shareable which, in turn, helps their collaborative and iterative advancement \cite{ackermann04}.

\begin{figure*}[!t]
     \centering

    \begin{subfigure}[t]{\miniminiteacherviewsize\textwidth}
        \centering
        \captionsetup{justification=centering}
        \includegraphics[height=240px,frame]{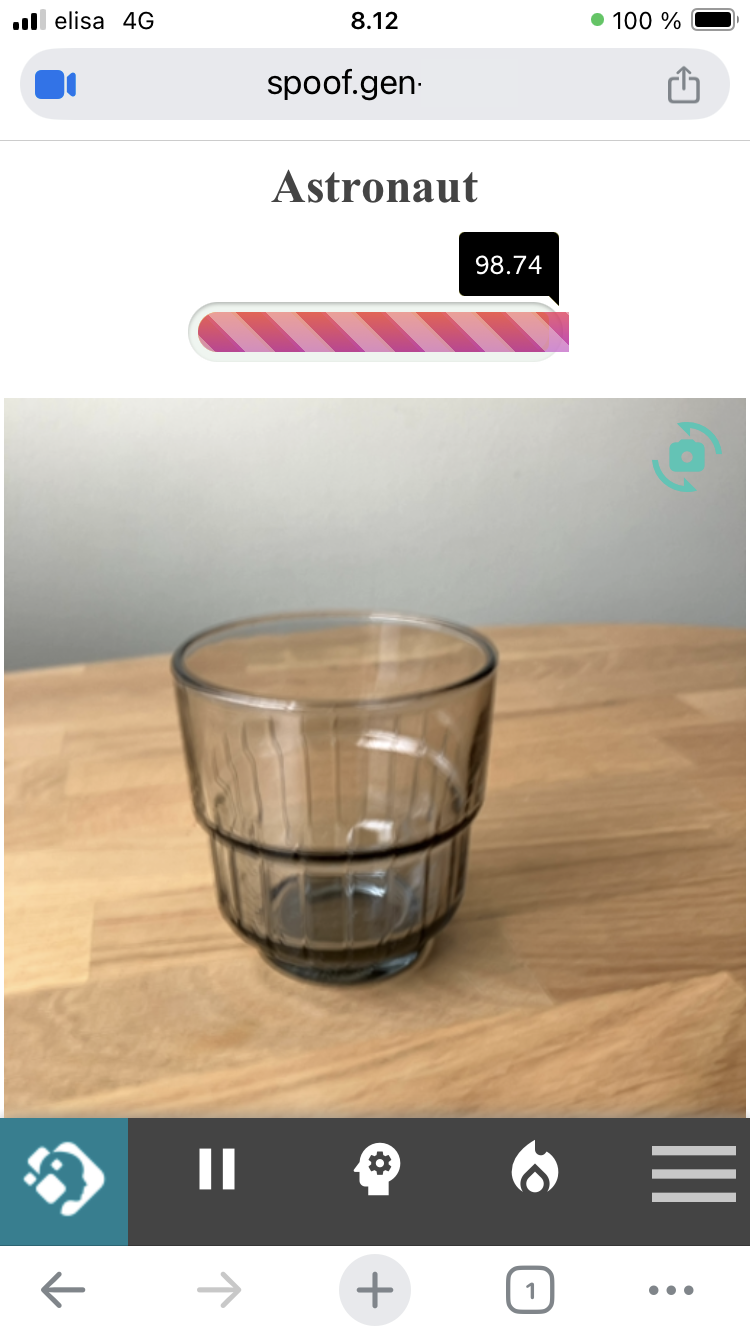}
        \caption{Spoofing app interface.}
        \label{fig:astronaut_interface}
    \end{subfigure}%
    ~ 
    \begin{subfigure}[t]{\miniminiteacherviewsize\textwidth}
        \centering
        \captionsetup{justification=centering}
        \includegraphics[height=240px,frame]{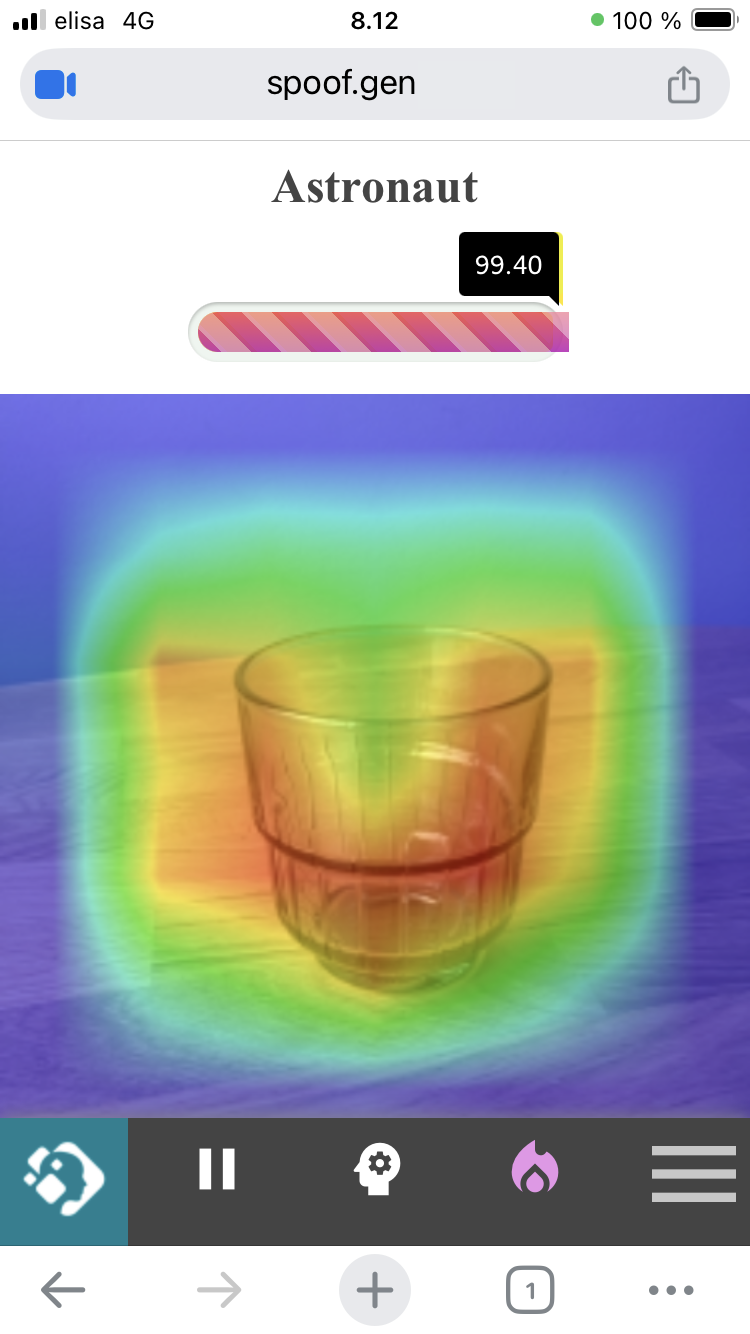}
        \caption{Heat map of salient image regions.}
        \label{fig:astronaut_heatmap}
    \end{subfigure}
    ~ 
    \begin{subfigure}[t]{\miniminiteacherviewsize\textwidth}
        \centering
        \captionsetup{justification=centering}
        \includegraphics[height=240px,frame]{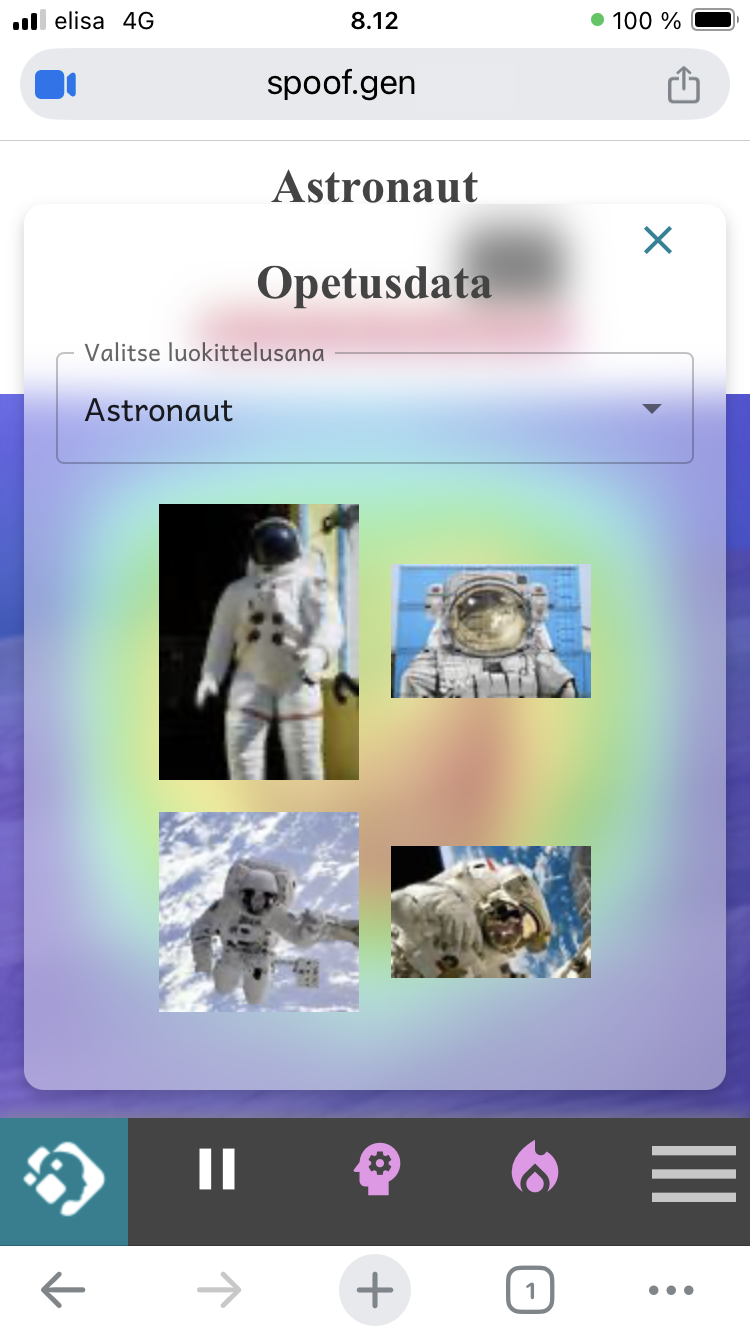}
        \caption{Examining the training data.}
        \label{fig:astronaut_trainingdata}
    \end{subfigure}%
    \caption{Example of spoofing the system with everyday items.  Figures \ref{fig:astronaut_interface} and \ref{fig:astronaut_heatmap} show an empty glass classified as ``astronaut'' with high confidence.  The training data view (Fig.\ \ref{fig:astronaut_trainingdata}) hints that the reason might be the astronaut's round glass visor, found in all images in the training data set.
    }
    \label{fig:misclassification}
\end{figure*}

This kind of exploration supports not only conceptual understanding but also what Freire \citeyearpar{freire70} termed as critical ``re-reading'' of the world.  Learners are not just passive users or observers of AI systems, but active agents who investigate and challenge, and even ``break'', them. By learning to manipulate the results of AI systems, students begin to see AI not as a black box but as an engineered product shaped by human decisions, assumptions, and biases.  This prepares them to grow as informed contributors in public debates and democratic decision-making about the use and governance of AI technologies. Understanding the limitations, brittleness, and fallibility of AI-driven systems is a critical component of AI literacy, particularly in an age where systems make high-stakes decisions and often convey a false sense of objectivity or certainty \cite{long20,tedre21}. 

Critically re-reading the human-engineered mechanims and societal impacts of AI system may also cultivate learners' volition and capacity to become transformative agents capable of questioning, resisting, reimagining, and rewriting existing sociotechnical systems and data-driven practices in pursuit of a more just and equitable AI-driven world.

\paragraph{AI concepts addressed.}

The game introduces students to key aspects of how image recognition systems operate and, more importantly, how they can succumb to adversarial attacks and fail.  It centers on four pedagogical activities: (1) analyzing how image recognition systems classify visual inputs, (2) exploring their fragility and failure modes, (3) experimenting with adversarial tactics to mislead an AI model, and (4) critically reflecting on the broader consequences of automated classification. These goals align with major AI education frameworks that emphasize the interplay of AI techniques, human-AI interaction, and societal impact \cite{miao24,touretzky23,long20}.  The intended learning outcomes (ILOs) focus on four core concepts related to AI and data.

\emph{Confidence and Correctness.}
The tool displays a confidence score, which represents the model's estimated probability that a given input belongs to a particular class, based on its internal parameters.  By allowing learners to experiment with ways to increase this score, the system helps them understand that a confident prediction is not necessarily a correct one  (see Fig.\ \ref{fig:misclassification}).  From the perspective of critical AI education, this emphasizes that machine ``certainty'' can be misleading, especially in manipulated situations.  

\emph{Adversarial Attacks (Spoofing).}
Students engage in hands-on adversarial play by modifying their appearance or environment to induce misclassification by the model, with high confidence.  Unlike synthetic adversarial attacks in the pixel space (see, e.g.,~\cite{papernot-adversarial-2017}), this game focuses on embodied, real-world spoofing that exposes vulnerabilities in the model's generalization.  This introduces learners to foundational issues with AI robustness and adversarial attacks, providing a concrete entry point into broader questions about AI safety and system reliability.  

\emph{Feature Sensitivity and Saliency.}  
Through iterative experimentation, students begin to recognize that image classifiers are highly sensitive to certain visual features---such as eyeglasses, headwear, skin tone, background objects, or lighting conditions---that may not be semantically meaningful. For example, why do automated job interview systems systematically give higher score to applicants with eyeglasses or with bookshelf in the background? \cite{narayanan24}  This introduces the concept of saliency: the system's strong weighting of certain input features in the classification outcome (e.g. the astronaut's visor in Fig.\ \ref{fig:misclassification}). By observing how seemingly insignificant visual changes can drastically shift outcomes, learners gain insight into how AI systems attend to data, and begin to ask not just what label is predicted, but what features the system is really responding to.

\emph{Brittleness and Failure.}
The activity facilitates understanding the vulnerability of AI based classification systems to sometimes unanticipated image features, challenging naïve conceptions of AI as smart, objective, stable, or infallible \cite{druga17}.  Through playful experimentation---changing props, poses, lighting conditions, clothing, or background---students observe how seemingly minor, semantically irrelevant, or intuitively wrong changes can significantly influence model outputs.  These interactions help to demystify AI and open space for critical discussions about generalization, robustness, reliability, and the limits of machine perception in real-world uses.

\emph{Misclassification and Its Real-World Consequences.} 
The activity encourages learners to move beyond surface-level experimentation toward interrogating the socio-technical dimensions of AI systems.  It raises questions like ``Why does the model respond this way?'', ``Who decides how the system is trained?'', \cite{gebru21,mitchell19} and ``What happens when AI systems misclassify people in real life?'' \cite{buolamwini18}.  By raising questions on misclassification, representational harm, and systemic bias, the game helps students recognize that AI systems are not neutral artifacts but products shaped by their design choices, training data, and cultural assumptions \cite{crawford21,eubanks18,birhane21}. Learners are encouraged to critically reflect on who defines model objectives, whose data are included or excluded, and who are most affected by misclassification, as well as imagine more just and accountable AI futures.

\emph{From Conceptual Understanding to Critical Participation in AI.} 
This game is designed not only to teach AI concepts but also to foster a mindset of inquiry, skepticism, and agency. By making visible the decision-making processes and vulnerabilities of image classifiers that are normally opaque to students it enables learners not only to become aware of these weaknesses but also to actively explore and manipulate them. By engaging with the fallibility of AI systems, students are positioned as critical investigators rather than passive users of technology---an important step toward critical participation in an AI-infused society. By that way, the game supports learners to develop a more critical understanding and attitude toward AI-driven systems, especially when embedded within purposeful pedagogical practices. Creative knowledge creation and critical inquiry mediated by this game can foster resistance and activism against the sociotechnical injustices that AI systems exacerbate, thereby providing essential scaffolding for cultivating transformative agency in the age of AI.


\section{Discussion}

While many existing AI education tools focus on training models and achieving functional outcomes \cite{carney20,pope25b}, this paper's approach taps on failure and breakdown. By encouraging students to manipulate a model and induce misclassification, the \textit{Breakable Machine} game creates opportunities for encountering and investigating expectation mismatches---moments when the system behaves in unexpected ways  \cite{dhanorkar21}. These moments of cognitive dissonance invite deeper inquiry into the inner mechanisms of AI, including the role of training data, feature saliency, and generalization. Rather than treating misclassifications as problems to be corrected, the game frames them as a possibility for exploitation, critical exploration, and epistemic investigation. This aligns with recent shifts in AI literacy research that emphasize not only technical understanding, but also epistemic and ethical interrogation of how AI systems operate and where they fail \cite{morales24,long20}. 

Central to our approach is the pedagogical re-framing of spoofing not as a glitch that can only hinder learning but as a tool for critical inquiry.  By intentionally tricking the system, students are exposed to its brittleness and manipulability, revealing how AI systems can produce outputs that are not only wrong but confidently so.  This hands-on adversarial play serves as an entry point into deeper questions about the sociotechnical complexity and fragility of ML systems, as well as safety and ethical accountability \cite{buolamwini18,crawford21}.

This game repositions learners from passive users or consumers to active interrogators. Rather than mastering an AI tool or adapting their behavior to better fit an AI system, students learn to challenge and manipulate them---what Freire \citeyearpar{freire70} might describe as ``re-reading the world.''  In that view, education is not about adapting to existing systems and oppressive practices, but about questioning the status quo, envisioning alternatives, and taking action for realizing that future. Similarly, it resonates with Stetsenko’s \citeyearpar{stetsenko19} notion of activist learning, which moves beyond reproduction and socialization into existing practices by positioning learners as active contributors and co-creators of social change.  In this light, spoofing becomes a pedagogical act of resistance that cultivates understanding and transformative agency.

By engaging in collective experimentation and observing their peers' spoofing strategies on the leaderboard, students co-construct knowledge about machine perception and its limits.  Students can observe, replicate, and refine each other's strategies, externalizing the reasoning and contributing to a shared understanding of AI behavior.  This social learning environment reveals how classifiers respond to real-world variation, and fosters epistemic agency as students learn to question what counts as ``correct'' in image recognition systems, and why. 


The game works with concepts that are less often addressed in K–12 contexts, such as adversarial attacks and feature saliency. While popular AI frameworks \cite{touretzky23,miao24,long20} emphasize perception, learning, and societal impact, few current tools specifically encourage learners to directly probe the limits and vulnerabilities of AI systems. Our approach introduces learners to these limitations not as incidental glitches, but as core content for critical reflection. By emphasizing how AI systems can be manipulated or misled, the game challenges overly deterministic or instrumentalist views of AI systems as neutral or robust. This complements the growing body of AI literacy work that emphasizes data agency, algorithmic accountability, and participatory ethics \cite[e.g.,][]{vartiainen24b,morales24,babai25}. It expands the curricular space of AI literacy by integrating critical inquiry, hands-on spoofing, and sociotechnical reflection.  Rather than shielding students from system flaws, we make them core content, preparing learners to navigate, critique, and shape AI-driven futures.

\section*{Acknowledgments}
Funded by the 
[omitted for review]
We extend our heartfelt thanks to the teachers and children who actively participated in developing and testing this material. 

\bibliography{bibliography}

\begin{thebibliography}{64}
\providecommand{\natexlab}[1]{#1}

\bibitem[{Abiodun et~al.(2020)Abiodun, Araujo, Balla, Harwood, Ruberto, Salgado, and Tang}]{abidoun20}
Abiodun, X.; Araujo, V.; Balla, V.; Harwood, Z.; Ruberto, D.; Salgado, B.; and Tang, A. 2020.
\newblock Erase Your Face.

\bibitem[{Ackermann(2004)}]{ackermann04}
Ackermann, E.~K. 2004.
\newblock Constructing Knowledge And Transforming the World.
\newblock In Tokoro, M.; and Steels, L., eds., \emph{A learning zone of one's own: Sharing representations and flow in collaborative learning environments}, 15--37. Amsterdam, The Netherlands: IOS Press.

\bibitem[{Aleknavi{\v c}i{\=u}t{\.e}, Lehtinen, and S{\"o}dervik(2023)}]{aleknaviciute23}
Aleknavi{\v c}i{\=u}t{\.e}, V.; Lehtinen, E.; and S{\"o}dervik, I. 2023.
\newblock Thirty years of conceptual change research in biology -- A review and meta-analysis of intervention studies.
\newblock \emph{Educational Research Review}, 41: 100556.

\bibitem[{Babai et~al.(2025)Babai, Gazulla, Iivari, and Kinnula}]{babai25}
Babai, N.; Gazulla, E.~D.; Iivari, N.; and Kinnula, M. 2025.
\newblock Navigating the Future of Data-Driven Systems: Children's perspectives on data and agency.
\newblock In \emph{Proceedings of the 24th Interaction Design and Children}, IDC '25, 124--139. New York, NY, USA: ACM.
\newblock ISBN 9798400714733.

\bibitem[{Birhane, Prabhu, and Kahembwe(2021)}]{birhane21}
Birhane, A.; Prabhu, V.~U.; and Kahembwe, E. 2021.
\newblock Multimodal datasets: misogyny, pornography, and malignant stereotypes.
\newblock \emph{arXiv.org}, (2110.01963).

\bibitem[{Black et~al.(2024)Black, George, Eguchi, Dempsey, Langran, Fraga, Brunvand, and Howard}]{black24}
Black, N.~B.; George, S.; Eguchi, A.; Dempsey, J.~C.; Langran, E.; Fraga, L.; Brunvand, S.; and Howard, N. 2024.
\newblock A Framework for Approaching {AI} Education in Educator Preparation Programs.
\newblock \emph{Proceedings of the AAAI Conference on Artificial Intelligence}, 38(21): 23069--23077.

\bibitem[{Buolamwini and Gebru(2018)}]{buolamwini18}
Buolamwini, J.; and Gebru, T. 2018.
\newblock Gender Shades: Intersectional Accuracy Disparities in Commercial Gender Classification.
\newblock In Friedler, S.~A.; and Wilson, C., eds., \emph{Proceedings of 1st Conference on Fairness, Accountability and Transparency}, 77--91.

\bibitem[{Cardenas, Molas, and Puertas(2023)}]{cardenas23}
Cardenas, M.-I.; Molas, L.; and Puertas, E. 2023.
\newblock Artificial Intelligence with {Micro:Bit} in the Classroom.
\newblock In Balogh, R.; Obdr{\v{z}}{\'a}lek, D.; and Christoforou, E., eds., \emph{Robotics in Education}, 337--350. Cham: Springer Nature Switzerland.
\newblock ISBN 978-3-031-38454-7.

\bibitem[{Carney et~al.(2020)Carney, Webster, Alvarado, Phillips, Howell, Griffith, Jongejan, Pitaru, and Chen}]{carney20}
Carney, M.; Webster, B.; Alvarado, I.; Phillips, K.; Howell, N.; Griffith, J.; Jongejan, J.; Pitaru, A.; and Chen, A. 2020.
\newblock Teachable Machine: Approachable Web-Based Tool for Exploring Machine Learning Classification.
\newblock In \emph{The 2020 CHI Conference on Human Factors in Computing Systems}, CHI EA '20, 1--8. New York, NY, USA: ACM.

\bibitem[{Crawford(2021)}]{crawford21}
Crawford, K. 2021.
\newblock \emph{Atlas of {AI}: {P}ower, Politics, and the Planetary Costs of Artificial Intelligence}.
\newblock New Haven, CT, USA: Yale University Press.

\bibitem[{Dhanorkar et~al.(2021)Dhanorkar, Wolf, Qian, Xu, Popa, and Li}]{dhanorkar21}
Dhanorkar, S.; Wolf, C.~T.; Qian, K.; Xu, A.; Popa, L.; and Li, Y. 2021.
\newblock Who needs to know what, when?: Broadening the Explainable {AI (XAI)} Design Space by Looking at Explanations Across the {AI} Lifecycle.
\newblock In \emph{Proceedings of the 2021 ACM Designing Interactive Systems Conference}, DIS '21, 1591--1602. New York, NY, USA: ACM.
\newblock ISBN 9781450384766.

\bibitem[{Druga, Otero, and Ko(2022)}]{druga22}
Druga, S.; Otero, N.; and Ko, A.~J. 2022.
\newblock The Landscape of Teaching Resources for {AI} Education.
\newblock In \emph{Proceedings of the 27th ACM Conference on on Innovation and Technology in Computer Science Education Vol. 1}, ITiCSE '22, 96--102. New York, NY, USA: ACM.

\bibitem[{Druga et~al.(2017)Druga, Williams, Breazeal, and Resnick}]{druga17}
Druga, S.; Williams, R.; Breazeal, C.; and Resnick, M. 2017.
\newblock ``{H}ey {G}oogle is It {OK} If {I} Eat You?'': Initial Explorations in Child-Agent Interaction.
\newblock In \emph{Proceedings of the 2017 Conference on Interaction Design and Children}, IDC'17, 595--600. New York, NY, USA: ACM.

\bibitem[{Eguchi et~al.(2025)Eguchi, Cottrell, Berg-Kirkpatrick, and de~Sa}]{eguchi25}
Eguchi, A.; Cottrell, G.; Berg-Kirkpatrick, T.; and de~Sa, V. 2025.
\newblock Research Experience for Teachers in the Interdisciplinary {AI} -- Report from the Summer {AI} Research Experience 2024.
\newblock In Hartshorne, R.; and Cohen, J., eds., \emph{Proceedings of Society for Information Technology \& Teacher Education International Conference 2025}, 1446--1453. Orlando, FL, USA: AACE.

\bibitem[{Eubanks(2018)}]{eubanks18}
Eubanks, V. 2018.
\newblock \emph{Automating Inequality: How High-Tech Tools Profile, Police, and Punish the Poor}.
\newblock New York, NY, USA: St. Martin's Press.

\bibitem[{Fagerlund, Palsa, and Mertala(2025)}]{fagerlund25}
Fagerlund, J.; Palsa, L.; and Mertala, P. 2025.
\newblock Exploration of domains of educational purpose in {K-12} data literacy education research.
\newblock \emph{Educational Research Review}, 46: 100663.

\bibitem[{Freire(1970)}]{freire70}
Freire, P. 1970.
\newblock \emph{Pedagogy of the Oppressed}.
\newblock New York, NY, USA: Continuum Publishing Group.

\bibitem[{Gebru et~al.(2021)Gebru, Morgenstern, Vecchione, Vaughan, Wallach, III, and Crawford}]{gebru21}
Gebru, T.; Morgenstern, J.; Vecchione, B.; Vaughan, J.~W.; Wallach, H.; III, H.~D.; and Crawford, K. 2021.
\newblock Datasheets for datasets.
\newblock \emph{Communications of the {ACM}}, 64(12): 86--92.

\bibitem[{Grover(2024)}]{grover24}
Grover, S. 2024.
\newblock Teaching {AI} to {K}-12 Learners: Lessons, Issues, and Guidance.
\newblock In \emph{Proceedings of {ACM} Computer Science Education ({SIGCSE}) 2024 Conference}, 1--7. Portland, OR, USA: ACM.

\bibitem[{Heintz and Roos(2021)}]{heintz21}
Heintz, F.; and Roos, T. 2021.
\newblock Elements Of {AI} - Teaching the Basics of {AI} to Everyone in {S}weden.
\newblock In \emph{Proceedings of the 13th International Conference on Education and New Learning Technologies (EDULEARN21)}, 2568--2572. Online: IATED.

\bibitem[{H{\"o}per and Schulte(2023)}]{hoper23}
H{\"o}per, L.; and Schulte, C. 2023.
\newblock The data awareness framework as part of data literacies in {K-12} education.
\newblock \emph{Information and Learning Sciences}.

\bibitem[{Iivari et~al.(2024)Iivari, Iversen, Smith, Schaper, Vent\"{a}-Olkkonen, Hartikainen, Sharma, Kinnula, Lehto, Holappa, and Molin-Juustila}]{iivari24b}
Iivari, N.; Iversen, O.~S.; Smith, R.~C.; Schaper, M.-M.; Vent\"{a}-Olkkonen, L.; Hartikainen, H.; Sharma, S.; Kinnula, M.; Lehto, E.; Holappa, J.; and Molin-Juustila, T. 2024.
\newblock Transformative agency -- the next step towards children's computational empowerment.
\newblock In \emph{Proceedings of the 23rd Annual ACM Interaction Design and Children Conference}, IDC '24, 322--337. New York, NY, USA: ACM.
\newblock ISBN 9798400704420.

\bibitem[{{Jatzlau} et~al.(2019){Jatzlau}, {Michaeli}, {Seegerer}, and {Romeike}}]{jatzlau19}
{Jatzlau}, S.; {Michaeli}, T.; {Seegerer}, S.; and {Romeike}, R. 2019.
\newblock It's not Magic After All -- Machine Learning in {S}nap! using Reinforcement Learning.
\newblock In \emph{2019 IEEE Blocks and Beyond Workshop (B\&B)}, 37--41. Memphis, TN, USA: IEEE.

\bibitem[{Kahila et~al.(2024{\natexlab{a}})Kahila, Vartiainen, Arkko, Lin, Pope, and Tedre}]{kahila24b}
Kahila, J.; Vartiainen, H.; Arkko, E.; Lin, A.; Pope, N.; and Tedre, M. 2024{\natexlab{a}}.
\newblock Enhancing Understanding of Data Traces and Profiling Among {K}--9 Students Through Interactive Classroom Game.
\newblock In \emph{Proceedings of The 19th WiPSCE Conference on Primary and Secondary Computing Education Research}, WiPSCE '24. New York, NY, USA: ACM.

\bibitem[{Kahila et~al.(2024{\natexlab{b}})Kahila, Vartiainen, Tedre, Arkko, Lin, Pope, Jormanainen, and Valtonen}]{kahila24}
Kahila, J.; Vartiainen, H.; Tedre, M.; Arkko, E.; Lin, A.; Pope, N.; Jormanainen, I.; and Valtonen, T. 2024{\natexlab{b}}.
\newblock Pedagogical framework for cultivating children's data agency and creative abilities in the age of {AI}.
\newblock \emph{Informatics in Education}, 23(2): 323--360.

\bibitem[{Lane(2021)}]{lane21}
Lane, D. 2021.
\newblock \emph{Machine Learning for Kids: A Project-Based Introduction to Artificial Intelligence}.
\newblock San Francisco, CA, USA: No Starch Press.

\bibitem[{Lee et~al.(2022)Lee, Zhang, Moore, Zhou, Perret, Cheng, Zheng, and Pu}]{lee22}
Lee, I.; Zhang, H.; Moore, K.; Zhou, X.; Perret, B.; Cheng, Y.; Zheng, R.; and Pu, G. 2022.
\newblock {AI} Book Club: An Innovative Professional Development Model for {AI} Education.
\newblock In \emph{Proceedings of the 53rd ACM Technical Symposium on Computer Science Education - Volume 1}, SIGCSE 2022, 202--208. New York, NY, USA: ACM.
\newblock ISBN 9781450390705.

\bibitem[{Lindner, Seegerer, and Romeike(2019)}]{lindner19}
Lindner, A.; Seegerer, S.; and Romeike, R. 2019.
\newblock Unplugged Activities in the Context of {AI}.
\newblock In Pozdniakov, S.~N.; and Dagien{\.{e}}, V., eds., \emph{Informatics in Schools. New Ideas in School Informatics}, 123--135. Cham: Springer International Publishing.
\newblock ISBN 978-3-030-33759-9.

\bibitem[{Long and Magerko(2020)}]{long20}
Long, D.; and Magerko, B. 2020.
\newblock What is {AI} Literacy? Competencies and Design Considerations.
\newblock In \emph{Proceedings of the 2020 CHI Conference on Human Factors in Computing Systems}, CHI '20, 1--16. New York, NY, USA: ACM.

\bibitem[{Long et~al.(2023)Long, Roberts, Magerko, Holstein, DiPaola, and Martin}]{long23}
Long, D.; Roberts, J.; Magerko, B.; Holstein, K.; DiPaola, D.; and Martin, F. 2023.
\newblock {AI} Literacy: Finding Common Threads between Education, Design, Policy, and Explainability.
\newblock In \emph{The 2023 CHI Conference on Human Factors in Computing Systems}, CHI EA '23. New York, NY, USA: ACM.
\newblock ISBN 9781450394222.

\bibitem[{Martins and Gresse Von~Wangenheim(2022)}]{martins22}
Martins, R.~M.; and Gresse Von~Wangenheim, C. 2022.
\newblock Findings on Teaching Machine Learning in High School: A Ten - Year Systematic Literature Review.
\newblock \emph{Informatics in Education}.

\bibitem[{Melsi\'{o}n et~al.(2021)Melsi\'{o}n, Torre, Vidal, and Leite}]{melsion21}
Melsi\'{o}n, G.~I.; Torre, I.; Vidal, E.; and Leite, I. 2021.
\newblock Using Explainability to Help Children Understand Gender Bias in {AI}.
\newblock In \emph{Proceedings of the 20th Annual ACM Interaction Design and Children Conference}, IDC '21, 87--99. New York, NY, USA: ACM.
\newblock ISBN 9781450384520.

\bibitem[{Miao and Shiohira(2024)}]{miao24}
Miao, F.; and Shiohira, K. 2024.
\newblock \emph{{AI} Competency Framework for Students}.
\newblock Paris, France: {UNESCO}.

\bibitem[{Mitchell et~al.(2019)Mitchell, Wu, Zaldivar, Barnes, Vasserman, Hutchinson, Spitzer, Raji, and Gebru}]{mitchell19}
Mitchell, M.; Wu, S.; Zaldivar, A.; Barnes, P.; Vasserman, L.; Hutchinson, B.; Spitzer, E.; Raji, I.~D.; and Gebru, T. 2019.
\newblock Model Cards for Model Reporting.
\newblock In \emph{Proceedings of the Conference on Fairness, Accountability, and Transparency}, FAT* '19, 220--229. New York, NY, USA: ACM.
\newblock ISBN 9781450361255.

\bibitem[{Morales-Navarro et~al.(2024{\natexlab{a}})Morales-Navarro, Kafai, Konda, and Metaxa}]{morales24b}
Morales-Navarro, L.; Kafai, Y.; Konda, V.; and Metaxa, D. 2024{\natexlab{a}}.
\newblock Youth as Peer Auditors: Engaging Teenagers with Algorithm Auditing of Machine Learning Applications.
\newblock In \emph{Proceedings of the 23rd Annual ACM Interaction Design and Children Conference}, IDC '24, 560--573. New York, NY, USA: ACM.
\newblock ISBN 9798400704420.

\bibitem[{Morales-Navarro and Kafai(2023)}]{morales23}
Morales-Navarro, L.; and Kafai, Y.~B. 2023.
\newblock Conceptualizing Approaches to Critical Computing Education: Inquiry, Design, and Reimagination.
\newblock In Apiola, M.; L{\'o}pez-Pernas, S.; and Saqr, M., eds., \emph{Past, Present and Future of Computing Education Research : A Global Perspective}, 521--538. Cham: Springer International Publishing.

\bibitem[{Morales-Navarro et~al.(2024{\natexlab{b}})Morales-Navarro, Kafai, Nguyen, DesPortes, Vacca, Matuk, Silander, Amato, Woods, Castro, Shaw, Akgun, Greenhow, and Garcia}]{morales24}
Morales-Navarro, L.; Kafai, Y.~B.; Nguyen, H.; DesPortes, K.; Vacca, R.; Matuk, C.; Silander, M.; Amato, A.; Woods, P.; Castro, F.; Shaw, M.; Akgun, S.; Greenhow, C.; and Garcia, A. 2024{\natexlab{b}}.
\newblock Learning about Data, Algorithms, and Algorithmic Justice on {TikTok} in Personally Meaningful Ways.
\newblock In \emph{Proceedings of the 18th International Conference of the Learning Sciences}, ICLS. International Society of the Learning Sciences.

\bibitem[{Morales-Navarro, Noh, and Kafai(2025)}]{morales25}
Morales-Navarro, L.; Noh, D.~J.; and Kafai, Y. 2025.
\newblock Building {babyGPTs}: Youth engaging in data practices and ethical considerations through the construction of generative language models.
\newblock In \emph{Proceedings of the 24th Interaction Design and Children}, 1021--1026. New York, NY, USA: ACM.
\newblock ISBN 9798400714733.

\bibitem[{Narayanan and Kapoor(2024)}]{narayanan24}
Narayanan, A.; and Kapoor, S. 2024.
\newblock \emph{AI Snake Oil: What Artificial Intelligence Can Do, What It Can't, And How to Tell the Difference}.
\newblock Princeton, NJ, USA: Princeton University Press.

\bibitem[{Olari and Romeike(2024)}]{olari24b}
Olari, V.; and Romeike, R. 2024.
\newblock Data-related concepts for artificial intelligence education in {K}--12.
\newblock \emph{Computers and Education Open}, 7: 100196.

\bibitem[{Olari et~al.(2024)Olari, Zoppke, Reger, Samoilova, Kandlhofer, Dagiene, Romeike, Lieckfeld, and Lucke}]{olari24}
Olari, V.; Zoppke, T.; Reger, M.; Samoilova, E.; Kandlhofer, M.; Dagiene, V.; Romeike, R.; Lieckfeld, A.~S.; and Lucke, U. 2024.
\newblock Introduction of Artificial Intelligence Literacy and Data Literacy in Computer Science Teacher Education.
\newblock In \emph{Proceedings of the 23rd Koli Calling International Conference on Computing Education Research}, Koli Calling '23. New York, NY, USA: ACM.

\bibitem[{Osterhaus et~al.(2021)Osterhaus, Brandone, Vosniadou, and Nicolopoulou}]{osterhaus21}
Osterhaus, C.; Brandone, A.~C.; Vosniadou, S.; and Nicolopoulou, A. 2021.
\newblock Editorial: The Emergence and Development of Scientific Thinking During the Early Years: Basic Processes and Supportive Contexts.
\newblock \emph{Frontiers in Psychology}, 12.

\bibitem[{Papernot et~al.(2017)Papernot, McDaniel, Goodfellow, Jha, Celik, and Swami}]{papernot-adversarial-2017}
Papernot, N.; McDaniel, P.; Goodfellow, I.; Jha, S.; Celik, Z.~B.; and Swami, A. 2017.
\newblock Practical Black-Box Attacks against Machine Learning.
\newblock In \emph{Proceedings of the 2017 ACM on Asia Conference on Computer and Communications Security}, 506–519. New York, NY, USA: Association for Computing Machinery.
\newblock ISBN 9781450349444.

\bibitem[{Pope et~al.(2025)Pope, Kahila, Vartiainen, and Tedre}]{pope25b}
Pope, N.; Kahila, J.; Vartiainen, H.; and Tedre, M. 2025.
\newblock Children's {AI} Design Platform for Making and Deploying {ML}-Driven Apps: Design, Testing, and Development.
\newblock \emph{{IEEE} Transactions on Learning Technology}, 18: 130--144.

\bibitem[{Rizvi, Waite, and Sentance(2023)}]{rizvi23}
Rizvi, S.; Waite, J.; and Sentance, S. 2023.
\newblock Artificial Intelligence teaching and learning in {K}-12 from 2019 to 2022: A systematic literature review.
\newblock \emph{Computers and Education: Artificial Intelligence}, 4: 100145.

\bibitem[{Sanusi et~al.(2022)Sanusi, Oyelere, Vartiainen, Suhonen, and Tukiainen}]{sanusi22}
Sanusi, I.~T.; Oyelere, S.~S.; Vartiainen, H.; Suhonen, J.; and Tukiainen, M. 2022.
\newblock A systematic review of teaching and learning machine learning in {K-12} education.
\newblock \emph{Education and Information Technologies}.

\bibitem[{Sentance and Waite(2022)}]{sentance22}
Sentance, S.; and Waite, J. 2022.
\newblock Perspectives on {AI} and data science education.
\newblock In \emph{AI, data science, and young people. Understanding computing education}, volume~3, 2--9. Cambridge, UK: Raspberry Pi Foundation.

\bibitem[{Shapiro and Fiebrink(2019)}]{shapiro19}
Shapiro, R.~B.; and Fiebrink, R. 2019.
\newblock Introduction to the Special Section: Launching an Agenda for Research on Learning Machine Learning.
\newblock \emph{{ACM} Transactions on Computing Education}, 19(4): 30:1--30:6.

\bibitem[{Smetana and Bell(2012)}]{smetana12}
Smetana, L.~K.; and Bell, R.~L. 2012.
\newblock Computer Simulations to Support Science Instruction and Learning: A critical review of the literature.
\newblock \emph{International Journal of Science Education}, 34(9): 1337--1370.

\bibitem[{Stetsenko(2019)}]{stetsenko19}
Stetsenko, A. 2019.
\newblock Radical-Transformative Agency: Continuities and Contrasts With Relational Agency and Implications for Education.
\newblock \emph{Frontiers in Education}, 4: 1--13.

\bibitem[{Tang et~al.(2019)Tang, Utsumi, , and Lao}]{tang19b}
Tang, D.; Utsumi, Y.; ; and Lao, N. 2019.
\newblock {PIC}: A Personal Image Classification Webtool for High School Students.
\newblock In \emph{International Joint Conference on Artificial Intelligence, {EduAI} Workshop}.

\bibitem[{Tedre, Denning, and Toivonen(2021)}]{tedre21b}
Tedre, M.; Denning, P.~J.; and Toivonen, T. 2021.
\newblock {CT 2.0}.
\newblock In \emph{21st Koli Calling International Conference on Computing Education Research}, Koli Calling '21, 1--8. New York, NY, USA: ACM.

\bibitem[{Tedre et~al.(2021)Tedre, Toivonen, Kahila, Vartiainen, Valtonen, Jormanainen, and Pears}]{tedre21}
Tedre, M.; Toivonen, T.; Kahila, J.; Vartiainen, H.; Valtonen, T.; Jormanainen, I.; and Pears, A. 2021.
\newblock Teaching Machine Learning in {K}--12 Classroom: Pedagogical and Technological Trajectories for Artificial Intelligence Education.
\newblock \emph{{IEEE} {A}ccess}, 9: 110558--110572.

\bibitem[{Touretzky et~al.(2025)Touretzky, Gardner-McCune, Hanna, Chen, and Pawar}]{touretzky25}
Touretzky, D.; Gardner-McCune, C.; Hanna, W.; Chen, A.; and Pawar, N. 2025.
\newblock Learning to Think like a Neuron in Middle School.
\newblock \emph{Proceedings of the AAAI Conference on Artificial Intelligence}, 39(28): 29212--29219.

\bibitem[{Touretzky, Gardner-McCune, and Seehorn(2023)}]{touretzky23}
Touretzky, D.; Gardner-McCune, C.; and Seehorn, D. 2023.
\newblock Machine Learning and the Five Big Ideas in {AI}.
\newblock \emph{International Journal of Artificial Intelligence in Education}, 33(2): 233--266.

\bibitem[{Trundle and Bell(2010)}]{trundle10}
Trundle, K.~C.; and Bell, R.~L. 2010.
\newblock The use of a computer simulation to promote conceptual change: A quasi-experimental study.
\newblock \emph{Computers \& Education}, 54(4): 1078--1088.

\bibitem[{Vartiainen et~al.(2024{\natexlab{a}})Vartiainen, Kahila, Tedre, L{\'o}pez-Pernas, and Pope}]{vartiainen24b}
Vartiainen, H.; Kahila, J.; Tedre, M.; L{\'o}pez-Pernas, S.; and Pope, N. 2024{\natexlab{a}}.
\newblock Enhancing children's understanding of algorithmic biases in and with text-to-image generative {AI}.
\newblock \emph{New Media \& Society}, 14614448241252820.

\bibitem[{Vartiainen et~al.(2024{\natexlab{b}})Vartiainen, Pellas, Kahila, Valtonen, and Tedre}]{vartiainen24c}
Vartiainen, H.; Pellas, L.; Kahila, J.; Valtonen, T.; and Tedre, M. 2024{\natexlab{b}}.
\newblock Pre-Service Teachers' Insights on Data Agency.
\newblock \emph{New Media \& Society}, 26(4): 1871--1890.

\bibitem[{Vartiainen and Tedre(2025)}]{vartiainen25e}
Vartiainen, H.; and Tedre, M. 2025.
\newblock The {CEDE} Model: A Learning-Sciences Based Approach for Critical and Transformative {K-12} {AI} Education.

\bibitem[{Veldhuis et~al.(2025)Veldhuis, Lo, Kenny, and Antle}]{veldhuis25}
Veldhuis, A.; Lo, P.~Y.; Kenny, S.; and Antle, A.~N. 2025.
\newblock Critical Artificial Intelligence literacy: A scoping review and framework synthesis.
\newblock \emph{International Journal of Child-Computer Interaction}, 43: 100708.

\bibitem[{Wang and An(2021)}]{wang21}
Wang, C.; and An, P. 2021.
\newblock Explainability via Interactivity? Supporting Nonexperts' Sensemaking of pre-trained {CNN} by Interacting with Their Daily Surroundings.
\newblock In \emph{Extended Abstracts of the 2021 Annual Symposium on Computer-Human Interaction in Play}, CHI PLAY '21, 274--279. New York, NY, USA: ACM.
\newblock ISBN 9781450383561.

\bibitem[{Wiatrek, Verma, and Martin(2025)}]{wiatrek25}
Wiatrek, N.; Verma, Y.; and Martin, F. 2025.
\newblock {Word2Vec4Kids}: Interactive Challenges to Introduce Middle School Students to Word Embeddings.
\newblock In \emph{Proceedings of the AAAI Conference on Artificial Intelligence}, volume~39, 29228--29235.

\bibitem[{Zhang, Lee, and Moore(2023)}]{zhang23}
Zhang, H.; Lee, I.; and Moore, K. 2023.
\newblock Preparing Teachers to Teach Artificial Intelligence in Classrooms: An Exploratory Study.
\newblock In \emph{Proceedings of 17th International Conference of the Learning Sciences (ICLS) 2023}, 974--977.

\bibitem[{Zhou et~al.(2016)Zhou, Khosla, Lapedriza, Oliva, and Torralba}]{zhou16}
Zhou, B.; Khosla, A.; Lapedriza, A.; Oliva, A.; and Torralba, A. 2016.
\newblock { Learning Deep Features for Discriminative Localization }.
\newblock In \emph{2016 IEEE Conference on Computer Vision and Pattern Recognition (CVPR)}, 2921--2929. Los Alamitos, CA, USA: IEEE Computer Society.

\end{thebibliography}

\end{document}